\documentclass[a4paper,11pt]{article}
\usepackage{pos}
\usepackage{dsfont}
\usepackage{hyperref}

\title{Overcoming exponential volume scaling in quantum simulations of lattice gauge theories}
\ShortTitle{Overcoming exponential volume gate count scaling for U(1) lattice gauge theories}

\author*[a]{Christopher F. Kane}
\author[b,c]{Dorota M. Grabowska}
\author[d]{Benjamin Nachman}
\author[d]{Christian W. Bauer}

\affiliation[a]{Department of Physics, University of Arizona, Tucson, AZ 85719, USA}
\affiliation[b]{InQubator for Quantum Simulation (IQuS), Department of Physics, University of Washington, Seattle, WA 98195, USA}
\affiliation[c]{Theoretical Physics Department, CERN, 1211 Geneva 23, Switzerland}
\affiliation[d]{Physics Division, Lawrence Berkeley National Laboratory, Berkeley, CA 94720, USA}

\emailAdd{cfkane@arizona.edu}

\abstract{Real-time evolution of quantum field theories using classical computers requires resources that scale exponentially with the number of lattice sites. Because of a fundamentally different computational strategy, quantum computers can in principle be used to perform detailed studies of these dynamics from first principles. Before performing such calculations, it is important to ensure that the quantum algorithms used do not have a cost that scales exponentially with the volume. In these proceedings, we present an interesting test case: a formulation of a compact U(1) gauge theory in 2+1 dimensions free of gauge redundancies. A naive implementation onto a quantum circuit has a gate count that scales exponentially with the volume. We discuss how to break this exponential scaling by performing an operator redefinition that reduces the non-locality of the Hamiltonian. While we study only one theory as a test case, it is possible that the exponential gate scaling will persist for formulations of other gauge theories, including non-Abelian theories in higher dimensions.}

\FullConference{%
The 39th International Symposium on Lattice Field Theory,\\
8th-13th August, 2022,\\
Rheinische Friedrich-Wilhelms-Universität Bonn, Bonn, Germany
}

\begin{document}
\maketitle
	
\section{Introduction}
Time evolution of quantum systems is one important class of problems where quantum computers are expected to have an exponential speed up relative to classical computers. One class of quantum field theories of particular interest are gauge theories, which are used to describe the physics of elementary particles, effective theories in nuclear physics, and condensed matter systems. Typically, gauge theories are studied using the Lagrangian formalism due to the fact that symmetries are manifest from this point of view. In order to study gauge theories using quantum computers however, one instead has to work with the Hamiltonian formalism. As with studying any physics problem computationally, it is of fundamental importance that the particular formulation chosen allows for efficient simulation using quantum computers.

One property of gauge theories that makes formulating the Hamiltonian non-trivial is that the Hilbert space of a given Hamiltonian spans multiple charge sectors. This fact can make quantum simulations of such a theory difficult for two reasons. Firstly, any noise from the environment during a quantum computation can cause a charge-violating transition and lead to incorrect results. Additionally, the naive basis of states for the Hilbert space is generally over-complete, requiring more qubits and gate operations than strictly necessary. Formulations have been developed that enforce gauge invariance including these unphysical states \cite{Halimeh_2020, Halimeh:2020ecg, Halimeh:2021lnv, Halimeh:2021vzf, Lamm_Scott_Yukari_2020, Tran_2021, PhysRevLett.109.125302, Banerjee_2012, https://doi.org/10.48550/arxiv.2012.08620, PhysRevX.3.041018, PhysRevA.90.042305, Stannigel_2014, Stryker_2019}, as well as removing them $\textit{a priori}$ \cite{Kaplan:2018vnj,Unmuth-Yockey:2018xak,Haase:2020kaj,Bender:2020ztu,PhysRevD.19.619,Bauer:2021gek}. Reviews of different approaches, including both analog and digital methods, can be found in Refs.~\cite{Wiese_2013, Zohar_2015, Dalmonte_2016, Aidelsburger_2021, Ba_uls_2020, Zohar_2021, Klco_2022, Bauer:2022hpo}.

In this work, we consider a formulation of pure U(1) gauge theory in 2+1 dimensions where unphysical states have been removed \textit{a priori} \cite{Bauer:2021gek}. We show that in this formulation, Suzuki-Trotter time evolution leads to an exponential volume scaling in the gate count, which is prohibitively expensive for realistic lattice sizes. We then show that this exponential volume scaling can be reduced to polynomial by performing a specific change of operator basis, and conclude by applying this change of basis to a specific lattice size.

\section{Gauge invariance and Gauss' Law}

The theory we consider is a pure U(1) gauge theory in 2+1 dimensions.  The Hamiltonian for the classical continuum theory is given by an integral over electric and magnetic fields
\begin{align}
    H = \int d^2x\, \left(\vec{E}(x)^2 + B(x)^2\right),
\end{align}
where the electric and magnetic fields are subject to their respective electric and magnetic Gauss' law, given by $\vec{\nabla} \cdot \vec{E}(x) = \rho(x)$ and $\vec{\nabla} \cdot B(x) = 0$, where $\rho(x)$ is the charge density. The quantum version of this theory is formulated on a finite lattice, and one has different options for what basis of states to use to represent the Hilbert space. Unless extra care is taken, the states in the Hilbert space of a given formulation do not automatically obey these Gauss law constraints. One example of such a formulation is the Kogut-Susskind formulation \cite{PhysRevD.11.395}. 

\begin{figure}
    \centering
    \includegraphics[width=0.5\textwidth]{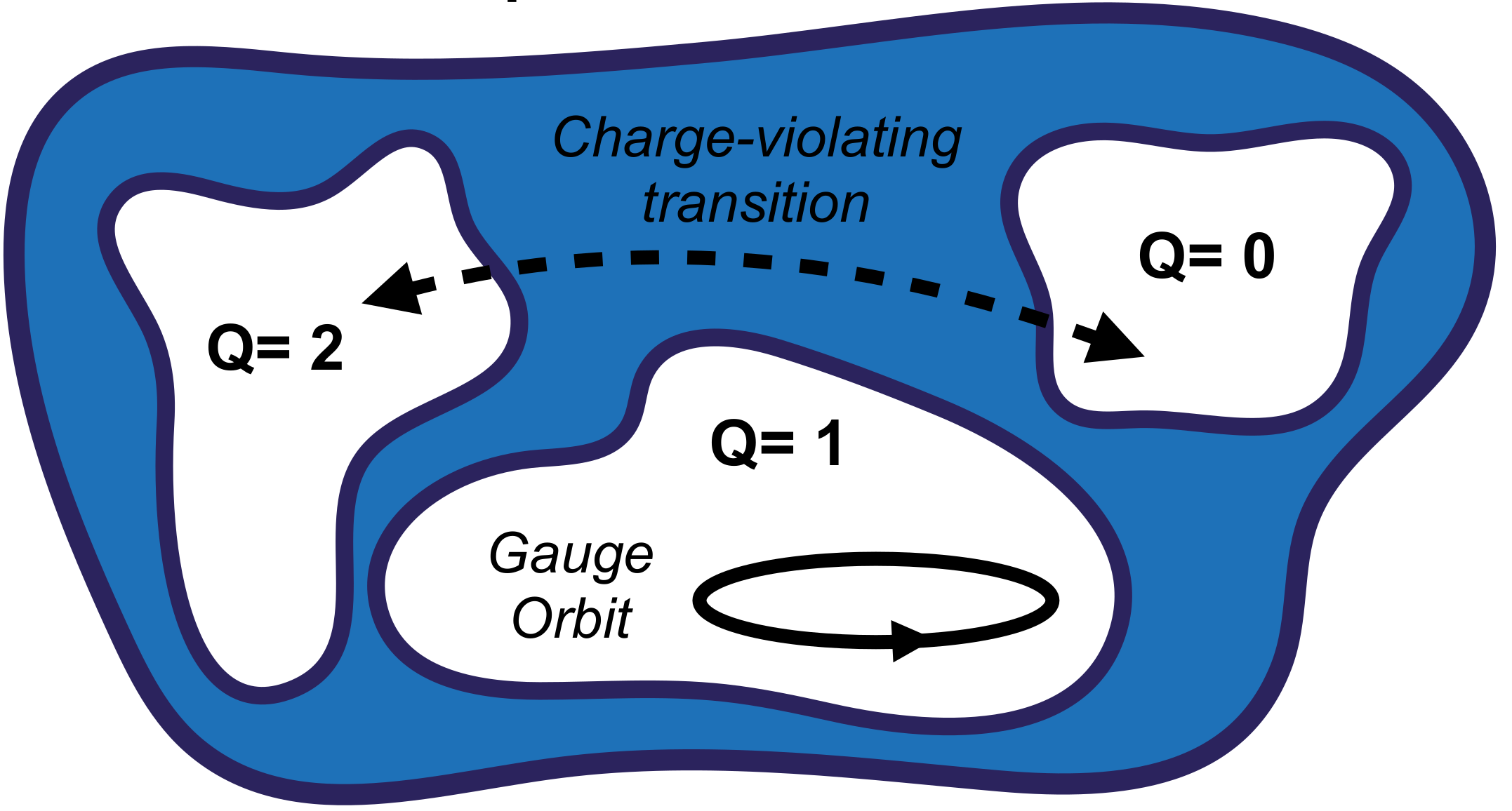}
    \caption{A schematic visualization of a Hilbert space. Different values of $Q$ represent different gauge sectors, each with a different total charge. A gauge orbit is the set of physically equivalent gauge fields in a particular charge sector.}
    \label{fig:hilbert_space}
\end{figure}

As mentioned in the introduction, this property leads to complications with regards to quantum simulations. Firstly, because this naive basis of states is over-complete due to the gauge orbit degeneracy, more quantum resources are required for the calculation than strictly necessary. Furthermore, because the Hilbert space spans multiple charge sectors, quantum noise in a quantum computation can cause unphysical transitions between them. Figure~\ref{fig:hilbert_space} shows a schematic visualization of a Hilbert space. When performing a quantum simulation, one first prepares an initial state in a specific charge sector, which can be thought of as a specific arrangement of static charges. In a noise-free quantum simulation, the state would evolve in time while remaining in the same initial charge sector. However, the presence of noise could cause an unphysical transition from the initial charge sector to a different charge sector. 
One solution to these problems is to work with a formulation of the U(1) theory whose Hilbert space is restricted to only physical states; in other words, a formulation that is free of gauge redundancies. Such a formulation would not allow charge-violating transitions, and could also require less quantum resources due to the reduced Hilbert space size. Much work has been done studying gauge-redundancy free formulations of U(1) gauge theories \cite{Kaplan:2018vnj,Unmuth-Yockey:2018xak,Haase:2020kaj,Bender:2020ztu,PhysRevD.19.619,Bauer:2021gek}, which in general have become known as ``dual basis'' formulations. 

The dual basis is formulated in terms of electric ``rotor'' operators and the usual magnetic field operators, both defined on the plaquettes of the lattice. The rotors are defined such that the transverse component of the electric field is given by $\vec{E}_T = \vec{\nabla} \times R$. By writing the longitudinal component of the electric field as $\vec{E}_L = \vec{\nabla} \cdot \rho$, one can see that the electric field $\vec{E} = \vec{E}_T + \vec{E}_L$ automatically satisfies electric Gauss' law. In these proceedings, we work in the charge-zero sector with $\rho(x)=0$. One important property of the rotor and magnetic field operators is that they are conjugate operators, satisfying $[B_p, R_{p'}] = i \delta_{pp'}$. This relation implies that the electric basis, where the rotor operators are diagonal, is related to the magnetic basis, where the magnetic field operators are diagonal, by the usual Fourier transformation. The Hamiltonian is given by a sum of the electric Hamiltonian $H_E$ and magnetic Hamiltonian $H_B$. For a periodic lattice with $N_x(N_y)$ sites in the $\hat{x}(\hat{y})$ directions, $H_E$ and $H_B$ are given by
\begin{equation}
    H_E = 2g^2 \sum_{p=1}^{N_x N_y} \left( \vec{\nabla} \times R_p \right)^2, \qquad H_B = \frac{1}{g^2} \left( N_x N_y - \sum_{p=1}^{N_x N_y} \cos B_p \right),
    \label{eq:HE_and_HB_pre_gauge_fix}
\end{equation}
where $g$ is the gauge coupling. If working in a sector with non-zero charge, the electric Hamiltonian will be modified appropriately. Note that we use lattice units throughout this work and therefore set the lattice spacing to one. 

While the local Gauss' law constraints are automatically satisfied, there is one remaining global constraint that is not automatically satisfied. This can be seen by counting the degrees of freedom in this formulation. Working in the magnetic basis, an eigenstate of the Hamiltonian in Eq.~\eqref{eq:HE_and_HB_pre_gauge_fix} is completely specified by $N_x N_y$ quantum numbers, one for each $B_p$ operator. However, it was shown in Ref.~\cite{Kaplan:2018vnj} that a gauge-fixed U(1) gauge theory in two spatial dimensions requires only $N_x N_y-1$ quantum numbers, which implies there is one redundant degree of freedom in the current formulation. This redundancy corresponds to the constraint that the product of plaquettes around a close surface must be the identity, which is the lattice version of the integral form of magnetic Gauss' law $\sum_{p=1}^{N_x N_y} B_p = 0$. In order to constrain the Hilbert space to contain only physical states, we choose to remove the redundancy by writing $B_{N_x N_y} = - \sum_{p=1}^{N_x N_y-1} B_p$, and setting the associated rotor to zero, \textit{i.e.} $R_{N_x N_y}=0$. In this way, the system now contains $N_p \equiv N_x N_y - 1$ independent plaquettes. While the new electric Hamiltonian has the same form as before, the magnetic Hamiltonian becomes highly non-local. In particular, the gauge-redundancy free magnetic Hamiltonian is given by (up to an overall constant)
\begin{equation}
    H_B = -\frac{1}{g^2} \left[ \sum_{p=1}^{N_p} \cos B_p + \cos \left(  \sum_{p=1}^{N_p} B_p \right) \right],
\end{equation}
where the term that is the cosine of the sum of all plaquettes $\cos(\sum_{p=1}^{N_p} B_p)$ is the source of the non-locality. We will show in Sec.~\ref{ssec:gateCount} that this cosine is the source of the exponential volume scaling in the gate count.

\section{Quantum simulation}
\label{sec:quantumSimulation}
In this section we describe the strategy used to perform the quantum simulation. We start by outlining the time evolution strategy used, and then review the digitization scheme of the $R_p$ and $B_p$ operators given in Ref.~\cite{Bauer:2021gek}. We conclude this section by studying how the gate count scales with the volume $N_p$. Note that we only provide the details of the digitization required to perform a gate count study. Further details can be found in Ref.~\cite{Bauer:2021gek}.

\subsection{Time evolution strategy}
\label{ssec:timeEvolutionStrategy}
To study the gate count, it is sufficient to consider first order Suzuki-Trotter methods. In particular, the time evolution operator is approximated as
\begin{equation}
    U(t) = \left( e^{-i \delta t H_E} e^{-i \delta t H_B} \right)^{N_\text{steps}} + \mathcal{O}(\delta t), 
\end{equation}
where $\delta t = t/N_\text{steps}$. We work in the magnetic basis, where $H_B$ is diagonal. Because rotors and magnetic field operators satisfy $[B_p, R_{p'}]=i \delta_{pp'}$, the electric and magnetic basis are related by a Fourier transform. Using this fact, the procedure we choose to perform a single Trotter step is to first implement the diagonal operator $e^{-i \delta t H_B}$, rotate to the electric basis using the quantum Fourier transform (QFT), implement diagonal operator $e^{-i \delta t H_E}$, and then rotate back to the magnetic basis using the inverse QFT. This procedure involves two classes of operators, namely the QFT, and diagonal operators. We perform a detailed gate count study of each step in Sec.~\ref{ssec:gateCount}.

\subsection{Digitization scheme}
\label{ssec:digitizationScheme}

The digitization scheme developed in Ref.~\cite{Bauer:2021gek} represents the rotors and magnetic field operators by diagonal matrices with evenly spaced eigenvalues. Each lattice site is represented by $n_q$ qubits, and so each operator is sampled $2^{n_q}$ times. Using the gate representation of such operators given in Ref.~\cite{PhysRevA.99.052335}, the $B$ and $R$ operators are written as\footnote{Note that Eq.~\eqref{eq:opRep} differs from Ref.~\cite{PhysRevA.99.052335} by a sign flip, a constant shift, and an overall multiplicative factor. This is due to a difference in the ordering of the eigenvalues, and the fact that we sample the operators asymmetrically about the zero eigenvalue.}
\begin{align}
    B = -\frac{b_\text{max}}{2^{n_q}} \left(\mathds{1} + \sum_{j=0}^{n_q-1} 2^j \sigma^z_j \right), \qquad R = -\frac{r_\text{max}}{2^{n_q}} \left(\mathds{1}+\sum_{j=0}^{n_q-1} 2^j \sigma^z_j\right),
    \label{eq:opRep}
\end{align}
where $\sigma^z_j$ is a Pauli-Z matrix acting on the $j^\text{th}$ qubit and $\mathds{1}$ is the $2^{n_q}\times 2^{n_q}$ identity matrix. The $B$ and $R$ operators are sampled in the ranges $[-b_\text{max}, b_\text{max}]$ and $[-r_\text{max}, r_\text{max}]$, respectively. Furthermore, the constant $b_\text{max}$ is chosen to minimize the digitization errors for a given value of the gauge coupling $g$, and the constant $r_\text{max}$ is a known function of $b_\text{max}$. Choosing a good value of $b_\text{max}$ can be done via a known analytic function and does not require performing a scan~\cite{Bauer:2021gek}. It was shown in Ref.~\cite{Bauer:2021gek} that sampling each operator a number of times corresponding to $n_q=3$ reproduces the low-lying spectrum to per-mille level accuracy. This fact will be important for the gate count study that follows.

\subsection{Gate count study}
\label{ssec:gateCount}

We are now in a position to study the gate count scaling for a single Trotter step. As previously mentioned, while $N_p$ will have to be taken large to simulate realistic lattice volumes, per-mill level accuracy can be achieved while keeping the number of qubits per lattice site $n_q$ small.

Each component of the Hamiltonian is implemented on a quantum computer separately, and we therefore consider the gate count of each component separately. Changing the basis requires performing a quantum Fourier transform. This can be done at each lattice site using a Fast Fourier Transform, which requires $O(n_q^2)$ gates per site \cite{nielsen_chuang_2010}. Therefore, the total gate count for changing the basis and back again for a single step is $O(n_q^2 N_p)$. Moving on to the electric Hamiltonian, because the lattice version of the curl is composed of differences, the electric Hamiltonian in Eq.~\eqref{eq:HE_and_HB_pre_gauge_fix} is composed of biliear terms of the form $R_p^2$ or $R_{p} R_{p'}$. Using the gate representation for $R$ in Eq.~\eqref{eq:opRep}, each binlinear term can be written schematically as $R^2 \sim \sum_{i,j=0}^{n_q-1} 2^{i+j} \sigma_i^z \sigma_j^z$, which is a sum of $n_q^2$ Pauli strings of length two. This implies that each bilinear term requires $\mathcal{O}(n_q^2)$ gates to exponentiate. The number of bilinear terms scales as $\mathcal{O}(N_p)$~\cite{Haase:2020kaj}, and so $e^{-i \delta t H_E}$ requires $\mathcal{O}(n_q^2 N_p)$ gates to implement. 

The number of gates required to implement the complex exponential of each $\cos(B)$ term in the magnetic Hamiltonian can be found by writing $B$ in terms of Pauli-Z operators and then expanding the cosine using the Taylor series. Doing so we find
\begin{equation}
    \cos(B) = \sum_{m=0}^{\infty} \frac{(-1)^m}{(2m)!} (B)^{2m} = \sum_{i_0=0}^1 \sum_{i_1=0}^1 \dots \sum_{i_{n_q-1}=0}^1 a_{i_0 i_1\dots i_{n_q-1}} (\sigma_0^z)^{i_0} (\sigma_1^z)^{i_1} \dots (\sigma_{n_q-1}^z)^{i_{n_q-1}},
    \label{eq:cosBPauliZ}
\end{equation}
where $a_{i_0 i_1\dots i_{n_q-1}}$ is the coefficient a given Pauli string and is in general non-zero. Equation~\eqref{eq:cosBPauliZ} indicates that each $\cos(B)$ term is a sum $2^{n_q}$ Pauli strings of length $n_q$, and therefore requires $\mathcal{O}(2^{n_q})$ gates to exponentiate. While this is exponential in $n_q$, recall that $n_q$ can be kept small and still reproduce the low-lying spectrum of the theory. More importantly, because there are $N_p$ such terms in $H_B$, exponentiation of all single cosine terms requires $\mathcal{O}(2^{n_q} N_p)$ gates, which is linear in the volume. Furthermore, because each $\cos(B)$ term acts on a different set of $n_q$ qubits, these terms can implemented in parallel, removing the runtime volume dependence. 

Following the same logic, using the Taylor series, the $\cos(\sum_{p=1}^{N_p} B_p)$ term is composed of a sum of $2^{n_q N_p}$ Pauli strings of length $n_q N_p$. The number of gates required to exponentiate this term is therefore $\mathcal{O}(2^{n_q N_p})$, which is exponential in the volume. For a modest sized $20 \times 20$ lattice, this term will require $\mathcal{O}(2^{400 n_q})$ gates per Trotter step.

\section{Operator basis change to break exponential volume scaling}
In the previous section we showed that the cosine of a single $B$ operator requires $\mathcal{O}(2^{n_q})$ gates to exponentiate, and that the cosine of the sum of every $B$ operator requires $\mathcal{O}(2^{n_q N_p})$ gates. This discussion generalizes such that the cosine of a sum of $M$ magnetic field operators requires $\mathcal{O}(2^{n_q M})$ gates to exponentiate. Therefore, an equivalent theory where any given cosine in the magnetic Hamiltonian contains a sum of no more than $M \sim \log_2 N_p$ magnetic field operators would require only a polynomial number of gates. In this section, we explain how this can be done using an operator change of basis put forth in Ref.~\cite{Grabowska:2022uos}. A detailed proof that such an operator basis change exists for any lattice volume, as well as an explicit efficient construction method, can be found in Ref.~\cite{Grabowska:2022uos}.

The strategy is to perform a change of operator basis
\begin{equation}
    B_p \to \mathcal{W}_{pp'} B_{p'}, \quad R_p \to \mathcal{W}_{pp'} R_{p'},
\end{equation}
where $\mathcal{W}$ is an orthogonal matrix. We will refer to this new operator basis as the weaved basis. Because $\mathcal{W}$ is orthogonal, the electric and magnetic basis are still related by a Fourier transform, and the time evolution strategy in the weaved basis is the same as outlined in Sec.~\ref{ssec:timeEvolutionStrategy}. The matrix $\mathcal{W}$ is chosen to be block diagonal with $N_s$ sub-blocks, written as $\mathcal{W} = \text{diag}(W_{d_{(1)}},  W_{d_{(2)}}, \dots,  W_{d_{(N_s)}})$. The matrices on the diagonal $W_d$ are so-called weaved matrices of dimension $d$. Because each $W_d$ is chosen to be orthogonal, the resulting matrix $\mathcal{W}$ is also orthogonal. 

Note that any basis change that reduces the number of magnetic field operators in the $\cos(\sum_{p=1}^{N_p} B_p)$ term will increase the number of magnetic field operators appearing in the local $\cos(B_p)$ terms. These effects must be balanced in such a way as to ensure no term contains more than $\mathcal{O}(\log_2 N_p)$ magnetic field operators. The properties of $\mathcal{W}$ that result in this are \cite{Grabowska:2022uos}
\begin{enumerate}
    \item $\mathcal{W}$ is block diagonal with $N_s \sim \log_2 N_p$ sub-blocks
    \item Each sub-block $W_d$ has dimension $d \sim N_p / \log_2 N_p$
    \item First column of any $W_d$ has entries all equal to $1/\sqrt{d}$
    \item Each row of $W_d$ has no more than $\lceil \log_2 d \rceil + 1$ non-zero entries.
\end{enumerate}
The first three properties reduce the number of terms that appear in the $\cos(\sum_{p=1}^{N_p} B_p)$ term from $N_p$ to $\mathcal{O}(\log_2 N_p)$ and therefore this term now requires $\mathcal{O}(N_p^{n_q})$ gates to implement. The last property ensures that each of the previously local cosine terms contain no more than $\mathcal{O}(\log_2(N_p/\log_2 N_p))$ operators, and exponentiation of one of these terms requires $\mathcal{O}(N_p/\log_2 N_p)^{n_q}$ gates. The number of gates required to implement the exponential of $H_B$ in the weaved basis is therefore
\begin{equation}
    {\rm Gates}(e^{-i \delta t H_B}) = \mathcal{O}(N_p^{n_q}) + \mathcal{O}(N_p(N_p/\log_2 N_p)^{n_q}).
\end{equation}
Note that the number of gates is now polynomial in the volume $N_p$. While the change of operator basis will introduce more terms in the electric Hamiltonian, because $H_E$ is bilinear, the maximum number of terms is $\mathcal{O}(N_p^2)$. The worse case scaling to exponentiate $H_E$ in the weaved basis is therefore quadratic in the volume, given by $\mathcal{O}(n_q^2 N_p^2)$.

We now work through an example of how to choose $\mathcal{W}$ for $N_p=16$. We first provide the matrix and then walk through each of the four conditions required. The matrix is given by
\begin{equation}
    \mathcal{W} = \begin{pmatrix}
        W_4 & 0 & 0 & 0 \\
        0 & W_4 & 0 & 0 \\
        0 & 0 & W_4 & 0 \\
        0 & 0 & 0 & W_4
    \end{pmatrix},
    \qquad W_4 =  \begin{pmatrix}
        \frac{1}{2} & -\frac{1}{\sqrt{2}} & -\frac{1}{2} & 0 \\
        \frac{1}{2} & \frac{1}{\sqrt{2}} & -\frac{1}{2} & 0 \\
        \frac{1}{2} & 0 & \frac{1}{2} & -\frac{1}{\sqrt{2}} \\
        \frac{1}{2} & 0 & \frac{1}{2} & \frac{1}{\sqrt{2}}
    \end{pmatrix}.
    \label{eq:weavedNp16}
\end{equation}
We see that the number of sub-blocks is $N_s = 4$, the dimension of each sub-block is $d=4$, and the first column of each sub-block is equal to $1/2$.  Additionally, each row of a given sub-block has no more than three non-zero entries. The weaved matrix in Eq.~\eqref{eq:weavedNp16} therefore satisfies the required properties, and the number of gates required to exponentiate the magnetic Hamiltonian in the weaved basis for $N_p=16$ is $\mathcal{O}(16^{n_q})$. 

\begin{figure}
\begin{minipage}{0.45\textwidth}
    \centering
    \includegraphics[width=0.9\textwidth]{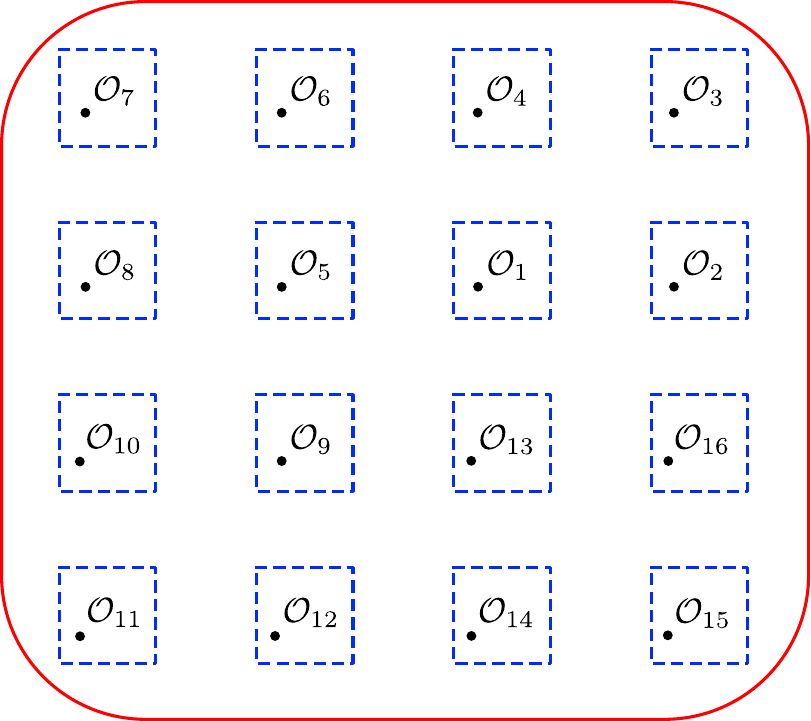}
\end{minipage}
\hfill\vline\hfill
\begin{minipage}{0.45\textwidth}
    \centering
    \includegraphics[width=0.9\textwidth]{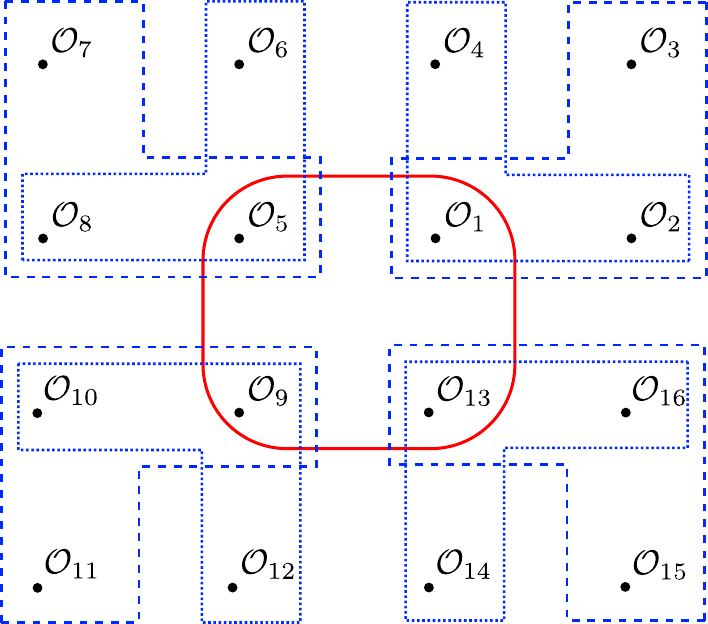}
\end{minipage}
\caption{Visual representation of the cosine terms in $H_B$ for $N_p=16$. Each operator $\mathcal{O}_p$ represents a magnetic field operator $B_p$, where the left(right) figure shows the original(weaved) basis. Operators inside the same box also appear in a single cosine of $H_B$. Boxes with different colors or line-styles correspond to different cosine terms. Left: The blue dashed rectangles with a single operator correspond to the local $\cos(B_p)$ terms in the original basis, and the red solid rectangle that contains every operator corresponds to the $\cos(\sum_{p=1}^{N_p} B_p)$ term in the original basis. Right: The blue dashed and dotted rectangles show, in the weaved basis, the increased connectivity of the previously local $\cos(B_p)$ terms. The red solid rectangle shows how the previously maximal connectivity of the $\cos(\sum_{p=1}^{N_p} B_p)$ term is reduced in the weaved basis.}
\label{fig:visual}
\end{figure}

This exponential reduction in gate count can also be understood visually. Figure~\ref{fig:visual} shows how the magnetic field operators are coupled in both the original and weaved basis, where the operators $\mathcal{O}_p$ represent the magnetic field operators, and all operators within the same box appear in a single cosine term in the Hamiltonian. The left image in Fig.~\ref{fig:visual} shows the connectivity in the original basis. The blue dashed boxes represent the $\cos(B)$ terms, and the red solid box represents the $\cos(\sum_{p=1}^{N_p} B_p)$ term. After performing the change of operator basis with the matrix $\mathcal{W}$ in Eq.\eqref{eq:weavedNp16}, the connectivity is given in the right image of Fig.~\ref{fig:visual}. The red solid box indicates that the number of magnetic field operators in the previously maximally coupled term has been reduced to four. The blue dashed and dotted boxes indicate that the previously local terms now each contain three magnetic field operators. The exponential reduction of the gate count in the weaved basis can be understood visually by the reduction in the maximum number of operators appearing in any given box from $N_p$ to $\log_2 N_p$. The number of gates required using \textit{e.g.} $n_q=2$ in the original basis is $\mathcal{O}(10^9)$ while in the weaved basis is only $\mathcal{O}(10^2)$.

\section{Conclusions}
Performing lattice gauge theory calculations on near-term quantum computers will require formulations that are robust to noise. It is also necessary, however, that the number of gates required to implement a given formulation does not scale exponentially with the system size. In these proceedings, we have shown that Suzuki-Trotter time evolution of a formulation of 2+1 U(1) lattice gauge theory that is fixed to a chosen charge sector, and therefore more robust to charge-violating transitions induced by quantum noise, requires a number of gates that scales exponentially with the volume. This exponential volume scaling will make even a single Trotter step for realistic lattice volumes prohibitively expensive. We demonstrated however, that this exponential volume scaling could be broken by performing an operator basis change to the weaved basis. This exponential reduction in the gate count will allow the possibility of simulating formulations of U(1) lattice gauge theories that are more robust to charge-violating transitions resulting from quantum noise.

Furthermore, while we applied the change of operator basis to a specific formulation of U(1) gauge theory in 2+1 dimensions, it can be applied to any bosonic theory. Moving forward, it will be interesting to see how it can be applied to formulations of non-abelian gauge theories.

\acknowledgments
The authors would like to thank Hank Lamm, Natalie Klco, Jesse Stryker and Tobias Osborne for useful discussions during the preparation of this work. CFK is supported by the U.S. Department of Energy (DOE) Computational Science Graduate Fellowship under award number DE-SC0020347. This work was supported by the DOE, Office of Science under contract DE-AC02-05CH11231, through Quantum Information Science Enabled Discovery
(QuantISED) for High Energy Physics (KA2401032), by the Office of Advanced Scientific Computing Research (ASCR) through the Accelerated Research for Quantum Computing Program, by the U.S. Department of Energy grant DE-FG02-97ER-41014 (Farrell), and the U.S. Department of Energy, Office of Science, Office of Nuclear Physics and \href{https://iqus.uw.edu}{\color{black}}{InQubator for Quantum Simulation (IQuS)} under Award Number DOE (NP) Award DE-SC0020970. This work is also supported, in part, through the \href{https://phys.washington.edu}{\color{black}}{Department of Physics} and \href{https://www.artsci.washington. edu}{\color{black}}{the College of Arts and Sciences}  at the University of Washington. This research benefited from the resources of the Oak Ridge Leadership Computing Facility, which is a DOE Office of Science User Facility supported under Contract DE-AC05-00OR22725.

\bibliography{references}
\bibliographystyle{ieeetr}
\end{document}